\documentclass[aps,prb,twocolumn,showpacs,preprintnumbers,amsmath,amssymb]{revtex4}

\usepackage{graphicx}
\usepackage{dcolumn}
\usepackage{bm}
\usepackage{color}
\usepackage{amsmath}

\begin{document}


\title{Quantum phase transition in a gapped Anderson model: NRG study }

\author{C. P. Moca$^{1,2}$  and A. Roman$^2$}
\affiliation{
$^1$Department of Theoretical Physics, Institute of
Physics, Budapest University of Technology and Economics, H-1521
Budapest, Hungary\\
$^2$Department of Physics, University of Oradea, Oradea, 410087, Romania
}

\date{\today}

\begin{abstract}
We use the numerical renormalization group method to investigate the 
spectral properties of a single-impurity Anderson model with a gap $\delta$ 
across the Fermi level in 
the conduction-electron spectrum. For any finite $\delta>0$, at half filling the 
ground state of the system is always a doublet. Away from half filling 
a quantum phase transition (QPT) occurs as function of the gap value $\delta$, 
and the system evolves from the strong-coupling (SC)
Kondo - type state, corresponding to $\delta < \delta_C$ toward a 
localized moment (LM) regime for $\delta > \delta_C$. 
The opening of the gap leads to the formation of one (two) bound states when 
the system is in the SC (LM) regime. The evolution across the QPT of their 
positions and the corresponding weights together with the dynamic properties of the model are 
investigated. 
\end{abstract}


\maketitle

\section{INTRODUCTION}

One of the hallmarks of the Kondo effect\cite{Hewson1993} is the raise 
of a narrow resonance, at the Fermi level,
in the spectral-density function of a magnetic 
impurity embedded into a metallic host. The 
width of the resonance is proportional to 
the so-called Kondo temperature $T_K$, which 
is the characteristic energy scale. Below 
$T_K$ the impurity spin is completely screened 
into a singlet by the host material. Below the Kondo scale ($T < T_K$), the low-temperature 
properties, such as the resistivity, spin susceptibility, or specific heat
are  properly described in terms of Landau theory\cite{Nozieres1974} of  the Fermi liquid.

The simplest approach to capture the Kondo effect is 
through the Anderson model.\cite{Anderson1961} It was 
used initially to describe the formation of the 
localized  moments (LMs) in metallic hosts. The model has inspired a lot of 
theoretical work, and a multitude of 
analytical and numerical methods was developed\cite{Hewson1993}. 
One of them, the numerical renormalization group
\cite{Wilson1975, Krishna-murthy1980_1, Krishna-murthy1980_2} (NRG), 
originally proposed by Wilson, is known as one of the most 
reliable and accurate approach to capture the 
low temperature, low-energy physics
of the model. Later, with the increase in the computing power, 
NRG was successfully extended 
to a broad range of more exotic quantum impurity models\cite{Bulla2008} such as 
the two-channel Kondo problem,\cite{Pustilnik2004} coupled magnetic
moments,\cite{Chung} the coupling to a superconducting host,\cite{Hecht2008}
or the soft-gap model.\cite{Bulla2000}

In the present work we address a slightly different problem. 
That of a magnetic impurity in a degenerate semiconductor  
host which presents a gap across the Fermi level in the 
conduction-electron spectrum. In the normal Anderson model the
conduction band has a flat density of states (DOS) at the Fermi level 
and the Kondo temperature is the only energy scale of the problem. 
The opening of a gap $\delta$ in the conduction-band spectrum introduces 
a new energy scale. The first question that arises is whether the 
Kondo state will survive? The problem  was originally 
addressed by using different techniques: the quantum Monte Carlo,\cite{Takegahara1992} 
density-matrix (DM) renormalization group,
\cite{Yu1996} Poor Man's scaling, and $1/N$ expansion \cite{Ogura1993} 
with no consensus reached. 
Later, Chen and Jayaprakash\cite{Chen1998} have used NRG method for the same problem.  
It was found that at 
half filling, any gap $\delta > 0$ changes the ground state to a doublet. 
Away from half filling and for 
large enough gaps the Kondo state does not survive.  
More recently, the results obtained within the NRG framework were confirmed by
using a local-moment approach.\cite{Galpin2008}  

The main goal of the present work is to extend the analysis of the model and to 
investigate its dynamic properties 
and their evolution across the quantum phase transition (QPT). In Sec. \ref{section:model}
we present the theoretical model, then the results for the spectral properties 
are presented in Sec \ref{section:spectral}. 
We give the conclusions in Sec \ref{section:conclusions}.

\section{MODEL AND NUMERICAL APPROACH}\label{section:model}

To describe a local, quantum impurity state, coupled to a conduction band we use the  
generic Anderson model

\begin{multline}
H = \sum_{\mathbf{k},\sigma}\epsilon_{\mathbf{k}}c_{\mathbf{k},\sigma}^{\dagger}c_{\mathbf{k},\sigma}+
\epsilon_d\,\sum_{\sigma}d_{\sigma}^{\dagger}d_{\sigma}\\
+U\,\sum_{\sigma}n_{d\uparrow}n_{d\downarrow}
+V\, \sum_{\mathbf{k},\sigma}\left(c_{\mathbf{k},\sigma}^{\dagger} d_{\sigma}+ 
d_{\sigma}^{\dagger}c_{\mathbf{k},\sigma}  \right)\label{eq:Hamiltonian}.
\end{multline}

Here $\epsilon_{\mathbf{k}}$ is the host band dispersion,
which is treated as a noninteracting one, 
$\epsilon_d$ is the impurity-level energy, $U$ is the on-site Coulomb energy 
at the impurity site, 
and $V$ is the hybridization-matrix element of the local impurity orbitals
with the band states,
which, in the present approach, is considered momentum and spin independent.
The number operator 
$n_{d\sigma}=d_{\sigma}^{\dagger}d_{\sigma}$, describes the occupation of the 
impurity level for spin-${\sigma}$ electrons. The mixing of impurity level with the 
host states is generically described by the hybridization function 
$\Delta\left( \omega \right) 
= \Delta_{\rm R }\left( \omega \right)+{\rm i}\, \Delta_{\rm I }\left( \omega \right)
=V^2\,\sum_{\mathbf {k}}\left[\omega -\epsilon_{\mathbf{k}} 
+{\rm i}\,\eta \,{\rm sgn}\left(\omega\right) \right]^{-1}$. In general 
$\Delta_{\rm I} (\omega)$ can be related to the density of states 
of the host band: $\Delta_{\rm I} (\omega)= \pi V^2 \varrho(\omega)$.  
In our model, a gap
$\delta>0$ is present in the density of states. Then, 
$\Delta_{\rm I} (\omega) $ has the form
\begin{equation}
\Delta_{\rm I}\left(\omega \right) = \Gamma\;
\Theta\left(\left| \omega\right|-\delta\right)
\Theta \left(D-\left| \omega\right|\right)
\label{eq:hybridization_imag_part}
\end{equation}
with $2D$ the bandwidth of the host band. 
The normal Anderson model with a flat density of states
is recovered in the limit of zero gap $\delta\rightarrow 0$, 
in which case the properly normalized density of states is
$\varrho (\omega)= 1/2D$ for $\omega \in [-D, D]$ and the 
broadening function becomes $\Gamma = \pi V^2/2D$. In 
the followings we will consider $D$ as the energy unit.
The real part of the hybridization function is obtained through the  Hilbert transform as
\begin{eqnarray}
\Delta_{\rm R}\left(\omega \right)& = & -\frac{\Gamma}{\pi}
\left[\;
\ln \left | \frac{\omega-D}{\omega-\delta}\right|
-\ln \left | \frac{\omega+D}{\omega+\delta}\right|  
\; \right]\nonumber\\
& \simeq&  -\frac{\Gamma}{\pi} 
\ln \left | \frac{\omega+\delta}{\omega-\delta}\right|, \;\;\;\;
\left| \omega \right |\ll D \label{eq:hybridization_real_part}.
\end{eqnarray}

The region of relevance corresponds to small gap values, of 
the order of Kondo temperature ( we use the Haldane's expression for 
the Kondo temperature,\cite{Haldane1978} so it is properly defined 
only in the limit of zero gap, $\delta =0$).
We are mostly interested in the dynamical properties of the model. 
These are best described by the single-particle, 
retarded Green's function of the impurity site: 
$
G_{ret}(t-t')= -{\rm i}\left<\left\{d_{\sigma}(t), 
d_{\sigma}^{\dagger}(t') \right\} \right> \Theta(t-t')
$.
In general, the 
time-ordered Green's
function of the interacting problem is given by the Dyson equation:
$G(\omega)= \left[ G_{0}(\omega)-\Sigma(\omega)\right]^{-1}$, 
in terms of the non-interacting, 
($U=0$)  Green's function 
$G_0(\omega)= \left[\omega-\epsilon_d-\Delta(\omega)\right]^{-1}$
and of the self-energy $\Sigma(\omega)$.  
Then, the spectral representation 
${\cal A} (\omega)=  -1/\pi\, \Im m\, G_{ret}(\omega) $ 
can be readily obtained. 

Away from the half filling it can be readily shown,\cite{Galpin2009} by 
the simple perturbation theory at the Hartree-Fock level, that 
outside the gap, ($\left | \omega \right | >\delta $) both 
real and imaginary parts of the self-energy are non-zero, 
so the spectrum is continuum. 
On the other hand inside the gap, ($\left| \omega\right| < \delta $)
the imaginary part of the self-energy vanishes.
Because of that, the $d$-level Green's function has a \emph{single} pole inside the gap. 
This pole corresponds to a resonant
state with a life-time inverse proportional with the
imaginary part of the self-energy. Therefore, it corresponds to  
a  \emph{real bound state} with infinite lifetime. The position of the bound 
state ($E_b$) can be obtained by solving the equation $G^{-1}_{ret}(E_b) = 0$ 
for $E_b$.
At the non-interacting level, (neglecting the self-energy corrections) 
the bound-state energy is determined from
\begin{equation}
E_b -\epsilon_d = \Delta_{\rm R}(E_b).\label{eq:bound_state}
\end{equation}
Away from the half-filling this 
equation has always only {\it one solution}. 
When the self-energy corrections are  included,
it only leads to a renormalization of the local energy 
$\tilde \epsilon_d= \epsilon_d + \Re e\, \Sigma (E_b)$, otherwise, 
the same scenario holds, and still a single bound state is formed
in the gap, but with the  position slightly shifted.

At half filling a more careful analysis is necessary. At the non-interacting
level we do expect a single bound state to appear exactly at zero energy. 
However, the same perturbative 
analysis gives for the imaginary part of the self-energy inside the gap:
$\Im m \Sigma(\omega) \propto \delta(\omega)$ and by the  Hilbert transform 
the real part becomes $\Re e \Sigma(\omega) \propto 1/\omega$. 
Using now Eq. \eqref{eq:bound_state}, but
with the self-energy correction included, it can be shown that at half filling,
a pair of two, particle-hole symmetric poles are formed inside the gap. 

The perturbative approach described above provides us with helpful information 
relative to the formation of the bound states. Regardless of its 
simplicity it has its own limitations. Computing the self-energy
in the second-order perturbation theory in $U$, the spectral function
of the impurity site can be resolved only at a qualitative level. More than that, in the 
large-$U$ limit the perturbation theory is supposed to fail. The previous 
analysis indicates that there is a fundamental difference between 
symmetric and asymmetric cases, but we can only speculate that 
the particle-symmetric case cannot be perturbatively connected to 
the non-interacting limit while the asymmetric model does. At the same time
no information relative to the 
nature of the ground state can be extracted. 
\begin{figure}[t]
  \includegraphics[width=0.9\columnwidth]{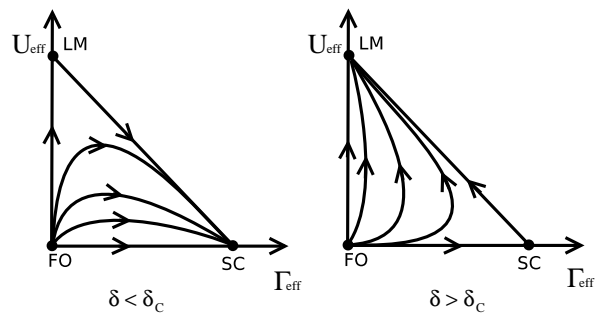}
  \caption{Renormalization group flow diagram at $T=0$
    for the gapped Anderson model, 
    when the system is away from half filling. When 
    $\delta< \delta_C$ the SC fixed point is stable while for 
    $\delta > \delta_C$ the LM becomes the stable fixed point. 
    At half filling $\delta_C =0$ so LM is stable irrespective 
    of the value $\delta >0$. }
  \label{fig:flow_diagram}
\end{figure}

A much more rigorous analysis of the problem is possible by using the NRG method,\cite{other}
 which is a reliable approach to study 
a variety of quantum impurity models. 
It  consists of a logarithmic discretization of the host band into 
intervals of the form 
$\left [\Lambda^{-(n+1)}, \Lambda^{-n} \right]$ with $\Lambda$
some positive number larger than 1, (usually $\Lambda\simeq 2 $) and a mapping of the 
original Hamiltonian (\ref{eq:Hamiltonian}) into a one-dimensional tight-binding
chain (Wilson chain) such that the hopping couplings between nearest neighbors
acquire a $\Lambda$ dependence of the form $\propto \Lambda ^{-n/2}$, 
followed by an iterative diagonalization of the chain with one extra site added
at each iteration step. 
In general  the on-site energies 
in the Wilson chain are also non-zero, but it can be analytically shown that
for a host density of states with electron-hole symmetry they do vanish. 
 The continuous limit corresponds to $\Lambda\rightarrow 1$	and for 
any $\Lambda > 1$ the NRG is an approximation. 	
Keeping $\Lambda$ small ($\simeq 1.5$) the computing time increases a lot while
having a large $\Lambda$ ($\sim 3.0$) the accuracy of the calculation is compromised,
especially for spectral properties at large energies.
Therefore, in the present work, we present results for $\Lambda=2$.
There is major difference in the way the NRG is used when a gap is opened in the
density of states.  For example, at $T=0$, in the normal Anderson model it is, in principle,
possible to increase the number of iterations to any value, but usually an upper limit 
is chosen such that, once the fixed point is reached,
 the NRG stops after a few iterations.
In principle, the more iterations we use, 
the better  the spectral functions at the Fermi level are resolved. 
On the other hand, for a gapped Anderson model the threshold
$\delta$ fixes somehow the maximum number of iterations. 
Because  there are no longer
states in the host band below $\delta$,  we need to stop the NRG procedure at a given 
iteration $N_{max}$ which is gap dependent: $N_{max} = N(\delta)$ 
such that the typical energy scale $\Lambda^{-(N_{max}-1)/2}$ {\it is not
much smaller} than the gap $\delta$. As a technical detail, 
we can take advantage
of the symmetries of the model and classify 
the  eigenstates of the Hamiltonian into multiplets. 
Here, we have used two quantum numbers to 
label the multiplets: $(i)$ $Q$ - the number of particle  
measured relative to the one particle per site;  $(ii)$ $S$ - the total spin.
In the normal Anderson model
the relevant energy scale is the Kondo temperature. It can 
be defined through the Haldane's expression:\cite{Haldane1978} 
$T^{0}_K = \sqrt{U \Gamma/ 2}\;e^{\pi \epsilon_d (\epsilon_d +U)/2U\,\Gamma}$. The 
gap $\delta$  introduces a new 
energy scale. As we will see next, the
relevant, dimensionless parameter that characterize 
the QPT is $\delta / T^0_K$.

\section{SPECTRAL FUNCTIONS AND THE PHASE DIAGRAM}\label{section:spectral}

\begin{figure}[t]
  \includegraphics[width=0.9\columnwidth]{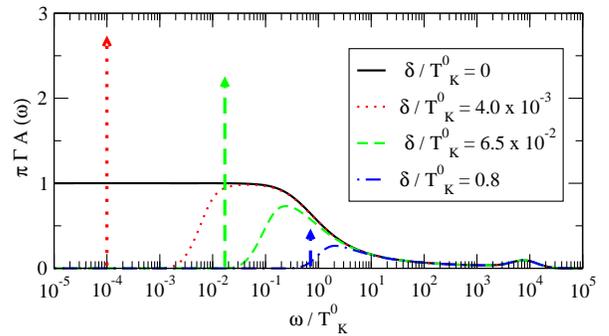}
  \caption{(Color online) Spectral function for the symmetrical model for different 
    gap values. The parameters used are $-\epsilon_d = \frac{U}{2} = 0.2$, $\Gamma =0.04$.
    $T^0_K$ is the Kondo temperature in the absence of the gap. The up arrows indicate
    the positions of the bound states in each case, their magnitude being proportional
    to the corresponding weight rescaled with the gap value, 
     i.e., $\pi \Gamma W_b / \delta$. Only the positive frequency 
    is plotted, the spectrum for negative energies being symmetric.}
  \label{fig:spectral_function_symmetric}
\end{figure}
\begin{figure}[t]
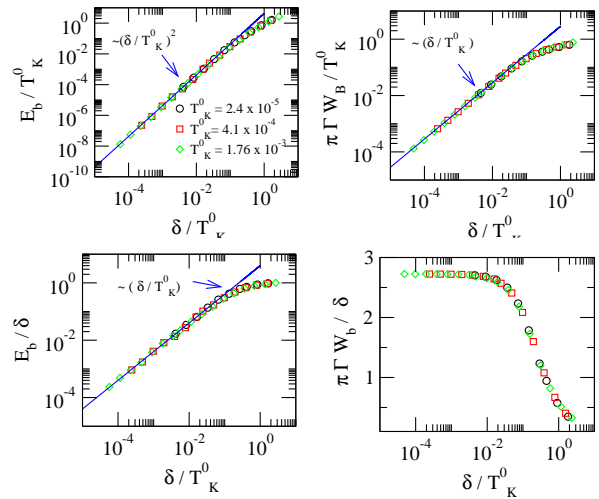

\begin{minipage}[b]{\linewidth}
\centering
\includegraphics[width=0.9\columnwidth]{peak_positions_symmetric_rescaled_with_Tk.eps}
\end{minipage}
\hspace{0.5cm}
\begin{minipage}[b]{\linewidth}
\centering
\includegraphics[width=0.9\columnwidth]{peak_positions_symmetric_rescaled_with_gap.eps}
\caption{ (Color online)  The positions of the bound states and 
  the corresponding spectral weights rescaled with the Kondo temperature $T^0_K$ (top layer) 
and with the gap value $\delta$ (bottom layer)} 
\label{fig:peak_positions_symmetric}
\end{minipage}
\end{figure}

We will start with a qualitative description of the 
renormalization group flow diagrams.  These are displayed 
in  Fig. \ref{fig:flow_diagram}.
We will consider first the symmetrical model. 
For a normal Anderson model the strong-coupling (SC) fixed
point is the only stable one, and the ground state is always a 
degenerate Kondo singlet characterized by the quantum numbers $(Q, S)=(\pm 1, 0)$. 
When a gap is opened in the density of states, the flow diagram completely changes. 
The SC fixed point
becomes unstable and the flow is toward the LM fixed point. 
At the same time the ground state changes to a doublet - $(0,\frac{1}{2} )$,
irrespective of the value of the gap, so for the symmetric model the critical gap
that describes the transition is $\delta_C = 0$. 

\begin{figure}[t]
  \includegraphics[width=0.9\columnwidth]{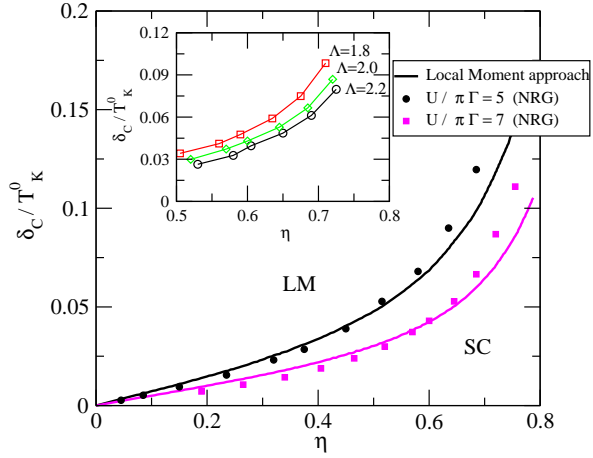}
  \caption{Phase diagram indicating the separation between SC and LM regimes. The parameter 
    $\eta = 1 + 2 \epsilon_d/U$  describes the asymmetry of the system. At half filling, the 
    critical gap is $\delta_C =0$,  and the system is always in the LM regime. The solid lines 
    correspond to the analytical results obtained within the 
    local-moment approach (see Ref. [\onlinecite{Galpin2008}])  with similar
     rescaled coupling $\tilde U =  U / \pi\, \Gamma $. The inset displays the 
     shift of the phase boundary line as function of $\Lambda$
     for $0.5<\eta<0.75$.}
  \label{fig:phase_diagram}
\end{figure}

Away from the half filling, and for the normal Anderson model ($\delta =0$),
probably the most interesting limit 
corresponds to the case when $\Gamma \ll -\epsilon_d \ll U $. 
Correspondingly, the flow is toward the frozen impurity (FI) fixed point, 
which on the other hand can be identified with the SC fixed point.\cite{ Krishna-murthy1980_1}
When a gap starts to open in the density of states the FI
remains stable as long as the gap is smaller than a critical value 
$\delta_C$. For gap values larger than $\delta_C$ ($\delta _C$ depends 
on the asymmetry of the problem, see Fig. \ref{fig:phase_diagram}) the FI becomes unstable
and the flow is toward the LM regime. At the same time the ground state
changes accordingly as in the case of symmetrical model from a singlet 
to a doublet.

In the following we will  present
results for the impurity spectral function, 
in these regimes.
In general, within the NRG framework, the spectral 
function is given as a weighted sum \cite{Bulla2008} of $\delta$ functions of the
form
\begin{equation}
{\cal A} (\omega)=  \sum_{i} W_i \delta\left ( \omega-\omega_i \right), 
\end{equation}
where the weights $W_i$ can be computed directly with the NRG. 
To get a smooth spectral function the $\delta$ functions need to be replaced
by some smooth kernels such that the spectral sum rule 
$\int d\omega {\cal A}(\omega)=1$ remains valid. 
When investigating the spectral properties of the gapped model 
we have used a slightly modified broadening procedure such that 
only the delta peaks which correspond to energies outside the gap 
($ \left | \omega  \right | > \delta $)  are 
broadened while for energies inside the gap  
($\left | \omega \right | < \delta $) the 
weights $W_b$ and the corresponding bound-state energies  $E_b$ are
extracted directly from the excitation spectrum, so the spectral
function becomes
\begin{equation}
{\cal A} (\omega)= \left . {\cal A}_{cont} (\omega)
\right |_{\left | \omega \right | > \delta}
+ \sum_{b}\left . W_b \delta\left ( \omega-E_b \right)
\right | _{\left | \omega \right | < \delta}.
\end{equation}

We will discuss separately the results for the cases when
the system is at/away from half filling. 
In  Fig. \ref{fig:spectral_function_symmetric} we present typical
results for the spectral function of the $d$-level of a symmetrical model.
The presence of the gap in DOS preserves the electron-hole symmetry, so the spectrum 
remains symmetric. The black solid line is the spectral function 
for the normal Anderson model which develops the usual Kondo resonance below $T^0_K$.
Below the Kondo temperature the localized spin 
is screened by the conduction electrons, and the ground 
state is a singlet corresponding to $S=0$.  
The opening of a gap in the density of states 
changes the physics dramatically. First of all, 
because no states are longer available at the Fermi level
below the gap edge $\delta$, the localized spin is no longer Kondo quenched 
into a singlet, and the ground state changes to a doublet $(0,\frac{1}{2})$. 

At the same time,
the  spectral function ${\cal A} (\omega)$ develops a gap and vanishes for energies
$\left | \omega \right | < \delta $ while \emph {two symmetric bound states} at $\pm E_b $  with
 exactly the same weights develop inside the gap  ($E_b < \delta$). Although the ground state
changes from a singlet to a doublet for any 
$\delta \ll T^0_K$ there are some reminiscence features of the 
Kondo peak and the behavior at energies $\left | \omega \right | > \delta $ 
resembles that of the normal Anderson model. For small enough gap values the bound states
are deep inside the gap. Increasing $\delta$, the $E_b$ moves toward the band edge. 
At the same time there is some transfer of spectral weight from the bound states to 
the continuum states. In the limit of $\delta \gg T^0_K$ the bound states merge with 
the continuum, their spectral weights becoming vanishingly small  
and it cannot be resolved any more. 
In Fig. \ref{fig:spectral_function_symmetric}  the positions of the 
bound states are indicated by up arrows while their magnitudes are rescaled with the 
gap value: $\pi\Gamma W_b /\delta$.
In Fig. \ref{fig:peak_positions_symmetric} the evolution of the rescaled bound-state energy and 
weights as function of the gap is plotted in two different ways: 
$(i)$ rescaled with $T^0_K$ - the Kondo temperature; $(ii)$
rescaled with $\delta$ - the value of the gap. 
We have found that 
the energies $E_b$ and the corresponding 
rescaled weights $w_b =\pi\, \Gamma\, W_b$ 
satisfy the following scaling equations in the limit   $\delta \ll T^0_K $:
\begin{eqnarray}
\frac{E_b}{T^0_K} & \propto & \left ( \frac{\delta}{T^0_K}\right )^2\nonumber\\
\frac{ w_b}{T^0_K} & \propto & \left (\frac{\delta}{T^0_K} \right).
\label{eq:scaling_equations}
\end{eqnarray}
\begin{figure}[t]
  \includegraphics[width=0.9\columnwidth]{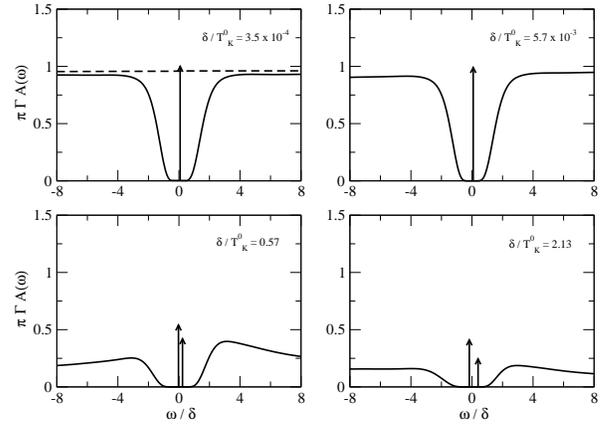}
  \caption{ Spectral functions for the  asymmetrical case, close to the Fermi energy. 
    In the upper panel
    the system is in the strong-coupling regime, characterized by $\delta \ll T^0_K$. 
    In the lower panel the system is in the local-moment regime with the
    gap $\delta $ of the order of $T^0_K$ or larger.  The arrows indicate the positions of the 
    bound states and their amplitude is rescaled to $\pi \Gamma W_b /\delta$. In the SC regime 
    a single resonance develops while in the LM regime there are two. 
    The dotted line in the left upper panel is the spectral function for the 
    normal Anderson model. }
\label{fig:spectral_function_asymmetric}
\end{figure}
\begin{figure}[t]
  \includegraphics[width=0.9\columnwidth]{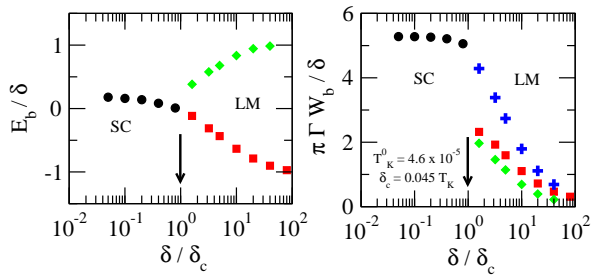}
  \caption{Typical evolution of the bound-state energies (left panel) and the corresponding weights
    (right panel) across the SC-LM quantum phase transition. In the left panel, 
    the circles (squares) represent the positions of the bound states in the SC(LM) regime.       
    The crosses in the right panel 
    indicate the sum of the weights of two bound states formed in the LM regime. The arrows
    point to the critical gap $\delta_C$ where transition occurs.}
  \label{fig:peak_positions_asymmetric}
\end{figure}
This scaling behavior can be understood in terms of an effective Hamiltonian.
First the Anderson Hamiltonian (\ref{eq:Hamiltonian}) is mapped by the help 
of the Schrieffer-Wolff transformations\cite{Schrieffer1996} into a Kondo problem and then a 
low-energy, effective Hamiltonian (at the energy scale $\delta$) is constructed, in which 
the impurity spin is coupled to the lowest electron/hole levels. The  Hamiltonian 
is described in terms of an effective exchange coupling $J_{eff} \propto T^0_K$ and 
a potential-like scattering term $K_{eff}\propto K_0\, T^0_K$. In the presence
of electron-hole symmetry, and in the LM regime, 
the scattering term vanishes: $K_{eff} =0$, while
only the exchange term survives. For a finite gap $\delta$
the system is always in a LM state with a ground state in the $(0, \frac{1}{2})$ sector 
while the first excited state is in the $(\pm 1, 0)$ sector. 
The energy $E_b$ of the bound state\cite{Chen1998}  
corresponds to the transition between the 
lowest energies within these sectors  $\left[\; (0, \frac{1}{2})\leftrightarrow (\pm 1, 0)\; \right]$ 
and  this energy difference scales as 
$E_b \simeq E_{(\pm 1,0)}- E_{(0, \frac{1}{2})} \propto \delta^2/T^0_K $. 

Next, we will describe the quantum phase transition when the system is away 
from the half filling. Here the SC regime is much robust as compared to the symmetric case
and the transition occurs always at a finite critical gap, $\delta_C$, such that for
$\delta<\delta_C$ the system is in the SC regime and for $\delta > \delta_C$ the 
flow is toward the LM fixed point. The critical gap $\delta_C$ depends on the  
 asymmetry parameter $\eta = 1+2\epsilon_d /U$.
In Fig. \ref{fig:phase_diagram} we present the separation between these 
two regimes in the parameter space $(\eta, \delta_C / T^0_K)$. We represent our numerical results 
for the critical values (symbols) together with the ones 
obtained by using the local-moment approach\cite{Galpin2008} (solid lines).  Close to the 
half filling the agreement between the methods is almost perfect,  and only 
for $\eta > 0.6-0.7$ deviations start to appear. 
The method of extracting the critical gap in the present approach is to 
some extend different from the one used in the soft-gap Anderson model \cite{Bulla2000}
where the phase transition boundaries can be extracted directly from the NRG flow diagram.
This kind of analysis is not possible in our approach since the NRG gets
truncated at the gap edge. In our approach the phase boundary
was obtained by changing the asymmetry parameter for a fixed gap and counting the number of
bound states in the gap. We believe that there is no other more efficient
procedure for constructing the phase diagram for this model.
More exactly, we have fixed the gap value $\delta$ as well as the Coulomb interaction $U$,
and we have changed the asymmetry parameter $\eta =1+2 \epsilon_d/U$. 
When the system is in the
SC regime, away from the half filling, 
there is always one bound state in
the gap, but decreasing the asymmetry, at some point
we cross the phase boundary and the system evolves toward the LM fixed point with two bound states in the gap. 
To give a quantitative description 
of the transition, 
in terms of the spectral properties, when the gap is changed smoothly from $\delta =0$ to
some large,  $\delta \gg T^0_K$, value,  we have focused on the regime with $U =0.4$, $\epsilon_d = -0.05$,
and $\Gamma = 0.01$. For this set of parameters the
Kondo temperature is $T^0_K = 4.6 \times 10 ^{-5}$ 
and the critical gap is found numerically to be $\delta_C = 0.045\, T^0_K$. 
In this particular regime (corresponding to $\Gamma \ll -\epsilon_d \ll U$ ) 
and for the normal Anderson model, 
below the Kondo temperature, the flow  is always toward the SC
fixed point, and the ground state is a non-degenerate singlet within the $(-1,0)$ channel. 
If we steadily increase $\delta$,
when $\delta \ll T^0_K $, the ground state remains a singlet. At the same time a gap is 
opening in the spectral function ${\cal A}(\omega)$ below  energies 
$\left | \omega \right | <\delta$. 
In the high-energy region, $ \left | \omega\right | > \delta $, 
${\cal A}(\omega)$ has a similar structure with that corresponding to the
normal Anderson model. The Kondo resonance forms below $T^0_K$ while 
the Hubbard side peaks develop at similar energies. 
In this SC regime only one bound state 
develops inside the gap (top layer in Fig. \ref{fig:spectral_function_asymmetric})
in agreement with the perturbation theory. 
As approaching the transition  $\delta \rightarrow \delta_C^-$ 
the energy of the bound state is shifted slowly toward the Fermi energy ($E_b \rightarrow 0$)
(see Fig. \ref{fig:peak_positions_asymmetric}). 
When $\delta $ becomes larger than $ \delta_C$ the SC is no longer the fixed point, 
and the flow is toward the LM regime. The ground state changes to a doublet: $(0, \frac{1}{2})$
while two non-symmetrical bound states develop in the gap. In Fig. \ref{fig:peak_positions_asymmetric}
we present the evolution of the bound-states energy together with their characteristic weights
across the transition point. On the LM side of the transition the localized states start to 
loose their weights as the gap is increased and their energy slowly merges into the continuum. 

\section{CONCLUSIONS}\label{section:conclusions}

Numerical renormalization group is by now a well-established method
for studying correlation effects in quantum impurity models. 
We have applied it here to investigate the  Anderson model 
with a gap in the conduction band. 
 By using a slightly modified version we were able
to capture the modifications in 
the spectral properties of the local operators 
across the quantum phase transition 
and the positions of the bound states formed inside the gap.
In this way we were able to construct the phase diagram of the model. 
The NRG provides one of the most accurate tools for investigating the dynamical 
quantities of quantum impurity models. When DM-NRG is used 
(such as in our case) the sum rules for the spectral
properties are satisfied up to the numerical precision. 
Therefore, the NRG results for the spectral functions 
are net superior to those obtained within other 
approaches such as the LMA  (see Ref. [\onlinecite{Galpin2008}] )  
which do capture correctly only
the positions of the bound states while the spectral properties 
suffer because of the approximations made.

\begin{acknowledgments}

This work has been supported by the Romanian
CNCSIS-UEFISCSU, under Grant No. PN II-IDEI 672/2009 and Hungarian OTKA 
under Grants No. NF061726 and No. K73361.

\end{acknowledgments}


\end{document}